\documentclass[final]{svjour2}
\usepackage{graphicx}
\usepackage{rotating}
\usepackage{amssymb}
\usepackage{mathptmx}
\usepackage[numbers]{natbib}
\makeatletter
\journalname{Journal of Low Temperature Physics}

\bibpunct{}{}{,}{s}{}{,}

\begin{document}

\newcommand{\hdblarrow}{H\makebox[0.9ex][l]{$\downdownarrows$}-}
\title{Development of Lumped Element Kinetic Inductance Detectors for the W-Band}

\author{A. Paiella$^{1,2,*}$ \and A. Coppolecchia$^{1,2}$ \and\\ M.G. Castellano$^{3}$ \and I. Colantoni$^{1}$ \and\\ A. Cruciani$^{1,2}$ \and A. D'Addabbo$^{4}$ \and\\ P. de Bernardis$^{1,2}$  \and S. Masi$^{1,2}$ \and G. Presta$^{1}$}

\institute{$^{1}$ Department of Physics, Sapienza, Rome, 00185, Italy\at$^{2}$ INFN-Sezione di Roma, Rome, 00185, Italy\at $^{3}$ CNR-IFN, Rome, 00100, Italy \at $^{4}$ Laboratori Nazionali del Gran Sasso (INFN), Assergi (AQ), 67100, Italy\\$^{*}$ \email{alessandro.paiella@roma1.infn.it}}

\maketitle

\begin{abstract}

We are developing a Lumped Element Kinetic Inductance Detector \\(LEKID) array able to operate in the W-band (75$-$110 GHz) in order to perform ground-based Cosmic Microwave Background (CMB) and mm-wave astronomical observations. The W-band is close to optimal in terms of contamination of the CMB from Galactic synchrotron, free-free, and thermal interstellar dust. In this band, the atmosphere has very good transparency, allowing interesting ground-based observations with large ($>$30 m) telescopes, achieving high angular resolution ($<$0.4 arcmin). In this work we describe the startup measurements devoted to the optimization of a W-band camera/spectrometer prototype for large aperture telescopes like the 64 m SRT (Sardinia Radio Telescope). In the process of selecting the best superconducting film for the LEKID, we characterized a 40 nm thick Aluminum 2-pixel array. We measured the minimum frequency able to break CPs (i.e. $h\nu=2\Delta\left(T_{c}\right)=3.5k_{B}T_{c}$) obtaining $\nu=95.5$ GHz, that corresponds to a critical temperature of 1.31 K. This is not suitable to cover the entire W-band. For an 80 nm layer the minimum frequency decreases to 93.2 GHz, which corresponds to a critical temperature of 1.28 K; this value is still suboptimal for W-band operation. Further increase of  the Al film thickness results in bad performance of the detector. We have thus considered a Titanium-Aluminum bi-layer (10 nm thick Ti $+$ 25 nm thick Al, already tested in other laboratories [1]), for which we measured a critical temperature of 820 mK and a cut-on frequency of 65 GHz: so this solution allows operation in the entire W-band.


\end{abstract}

\section{Science}

Millimiter-wave astronomy offers a variety of important scientific targets: from line emission of interstellar molecules to continuum emission of the CMB.

The W-band lies at an interesting transition between frequencies where Galactic emission is dominated by free-free and synchrotron, and frequencies where is dominated by dust (see : Fig. 1, [2]). In this band, the atmosphere is quite transparent, and then we can perform ground-based observations, avoiding the costs, complications and size limitations of balloon-borne and space-based missions.

\begin{figure}[!h]
\vspace{-0.4cm}
\begin{center}
\includegraphics[scale=0.15]{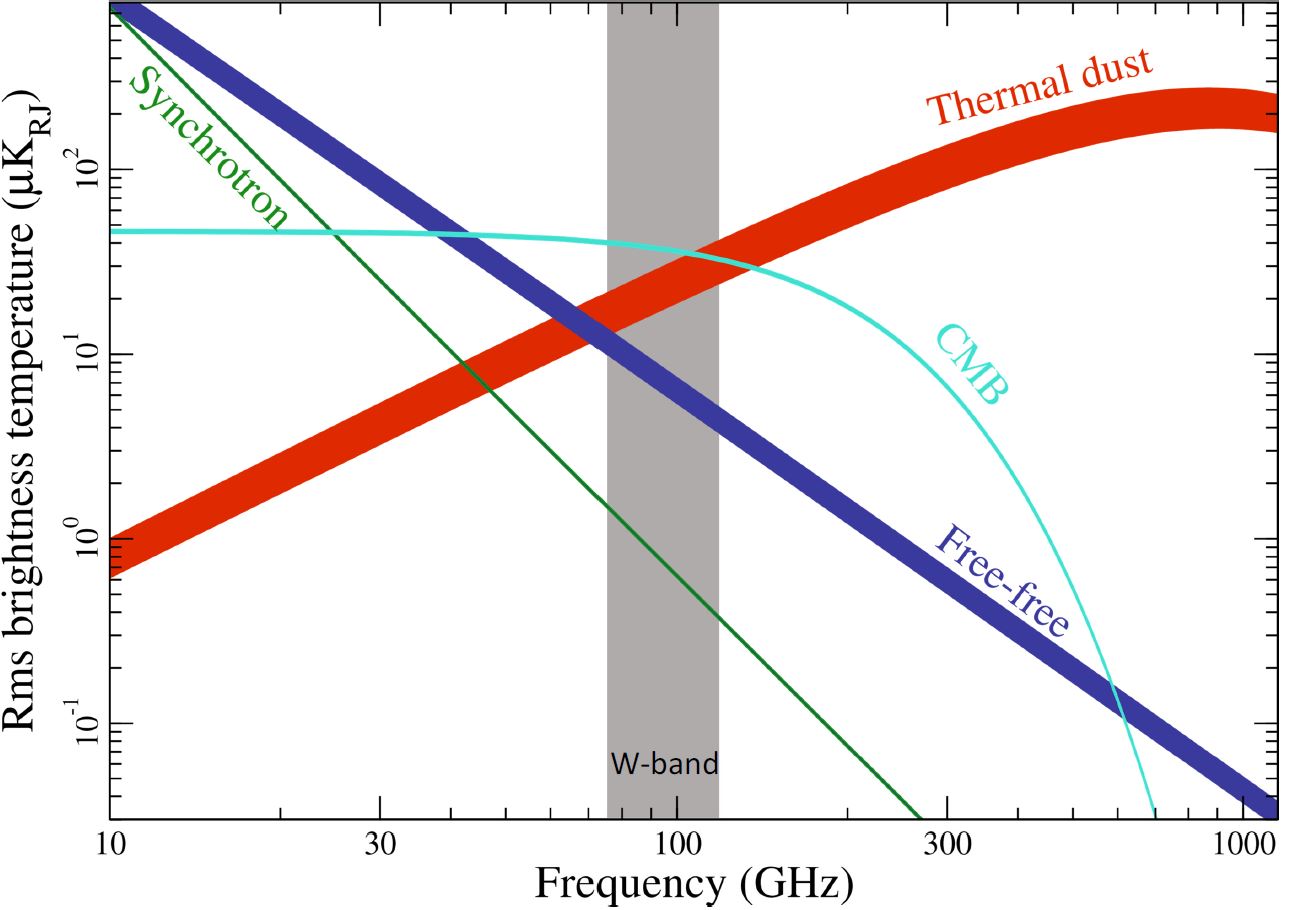}   
\end{center}
\vspace{-0.4cm}
\caption{Brightness temperature rms as a function of frequency for the astrophysical components. (Color figure online)}
\vspace{-0.4cm}
\label{fig:fig1}
\end{figure}

The measurements of B-modes of CMB polarization would provide the final confirmation of the cosmic inflation hypothesis [3]. To date this measurement has not been possible for two reasons: the sensitivity of the detectors and the polarized foreground emissions. The former can be improved with arrays of thousands of independent ultra-sensitive low-temperature detectors. The latter can be carefully separated from the interesting signal with multi-band experiments, where the W-band plays a key role. The same is true at smaller angular scales, where the Sunyaev Zeldovich effect of CMB photons crossing clusters of galaxies allows the study of the intracluster gas and even the use of clusters as cosmology probes [4].

\begin{figure}[!h]
\vspace{-0.6cm}
\begin{center}
\includegraphics[scale=0.30]{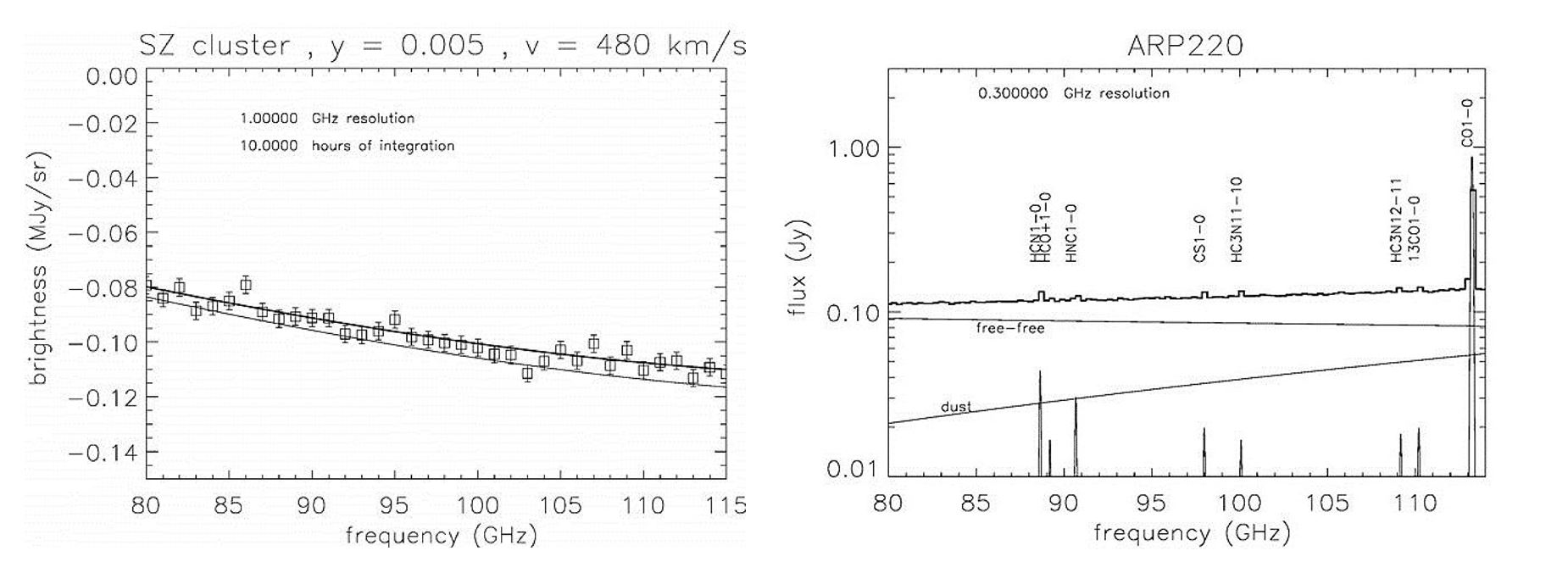}
\end{center}
\vspace{-0.4cm}
\caption{Simulations of the performance of a W-band FTS equipped with a 100 pixels array at the focus of the SRT. The {\it Left} panel shows a simulation of the measurement of the kinetic SZ effect, in the W-band, of a cluster with Comptonization parameter $y=0.005$ and peculiar velocity $v=480$ km/s, observed with a 1 GHz resolution FTS in 10 hours of integration. The {\it Right} panel shows a simulation of the molecular lines, in the W-band, for the merger galaxy ARP220.}

\label{fig:fig2}
\end{figure}

In Fig.~\ref{fig:fig2} we show two simulations of the performance of a W-band Fourier Transform Spectrometer (FTS) equipped with a 100 pixels array at the focus of the SRT.

\section{Experimental Setup}
Our cryogenic system is a four-stages cryostat [5], composed of a pulse-tube (PT) cryocooler, two absorption cryo-pumps (an $^{4}$He and an $^{3}$He
fridges) and an $^{3}$He$/^{4}$He dilution refrigerator. This single-shot cryostat is able to cool the detectors down to 180 mK for 7 hours.

\begin{figure}[!h]
\begin{center}
\vspace{-0.5cm}
\includegraphics[scale=0.135]{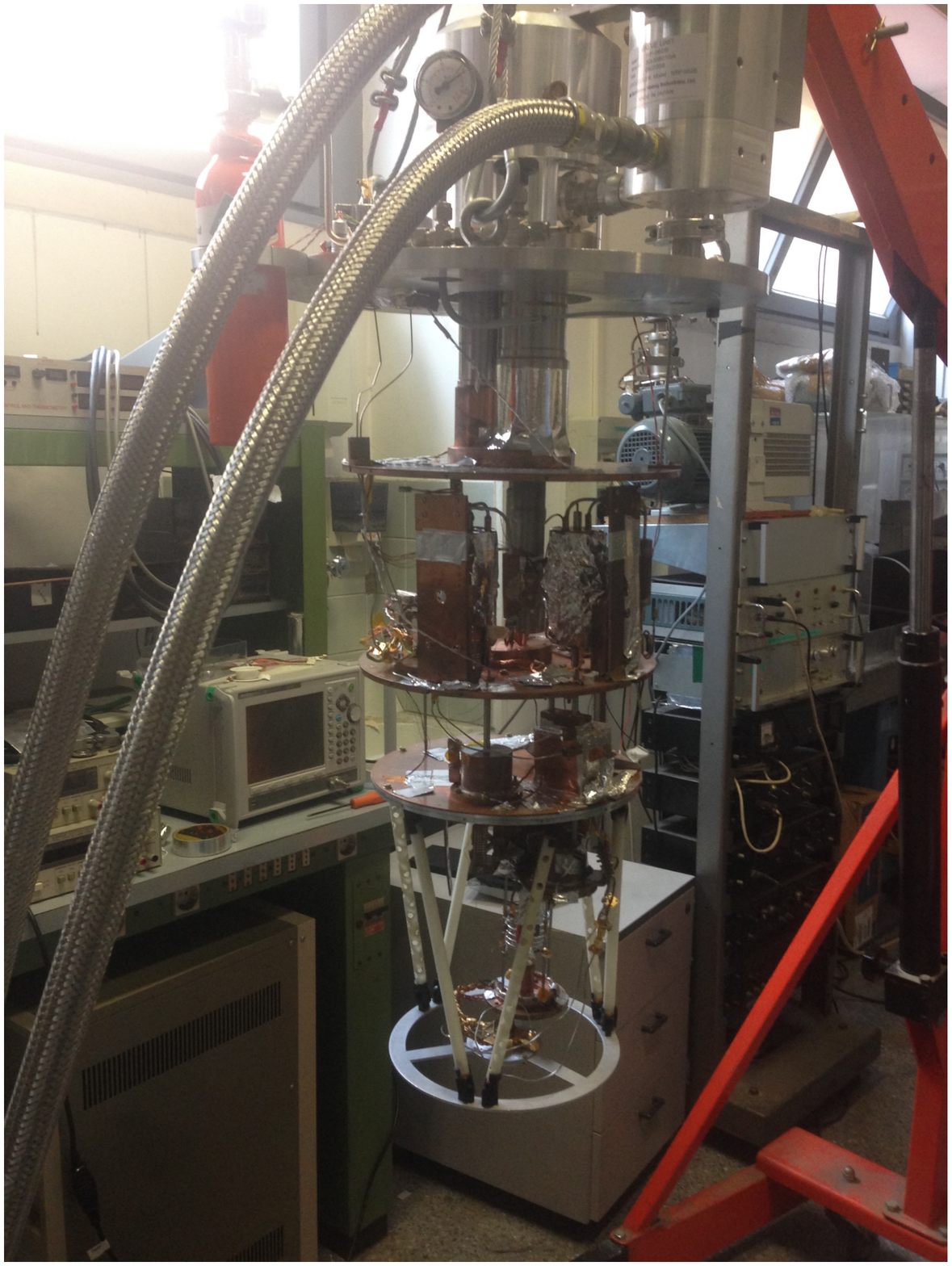}\hspace{-0.2cm}             
\includegraphics[scale=0.15]{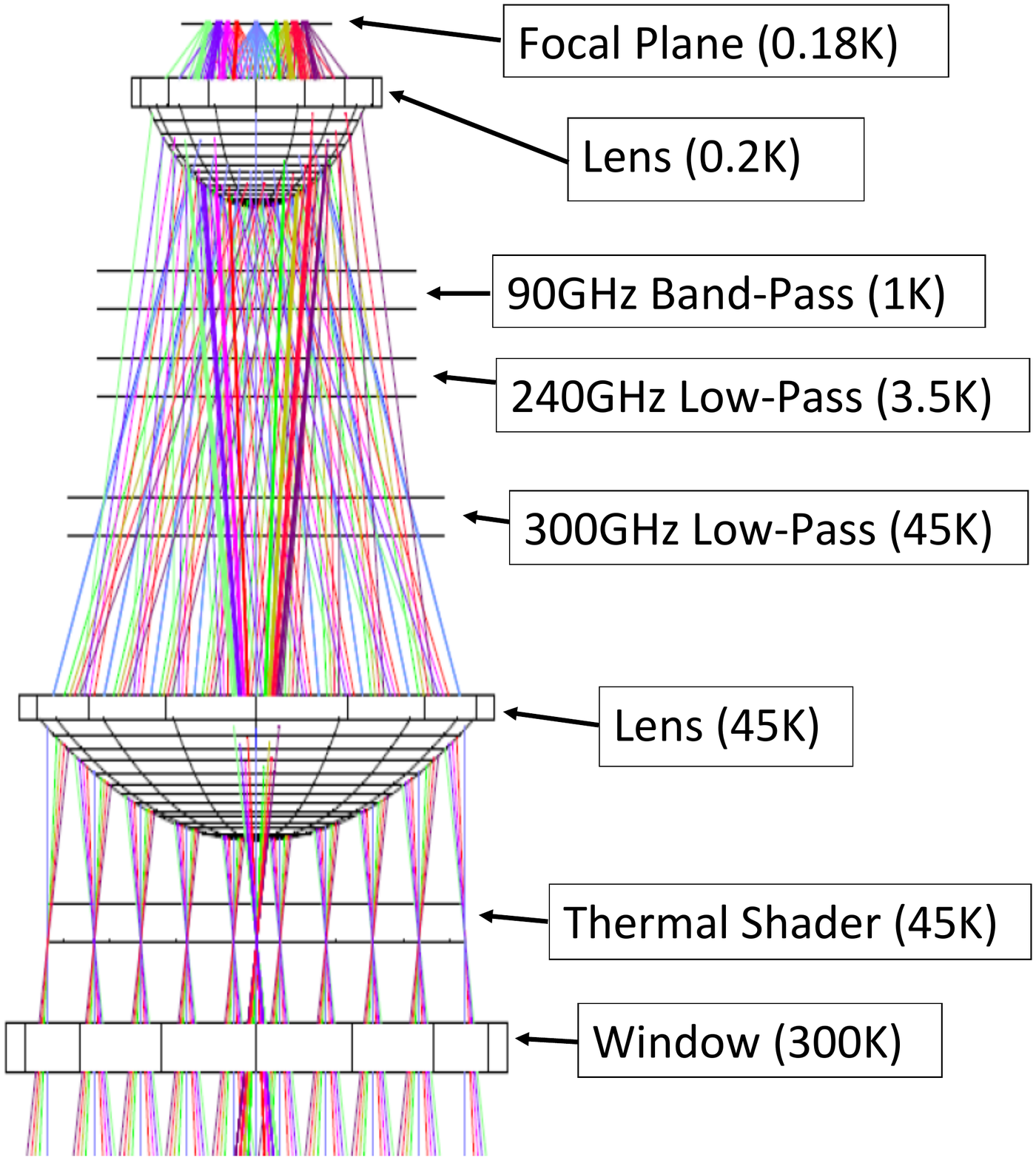}\hspace{-0.35cm}                
\includegraphics[scale=0.34]{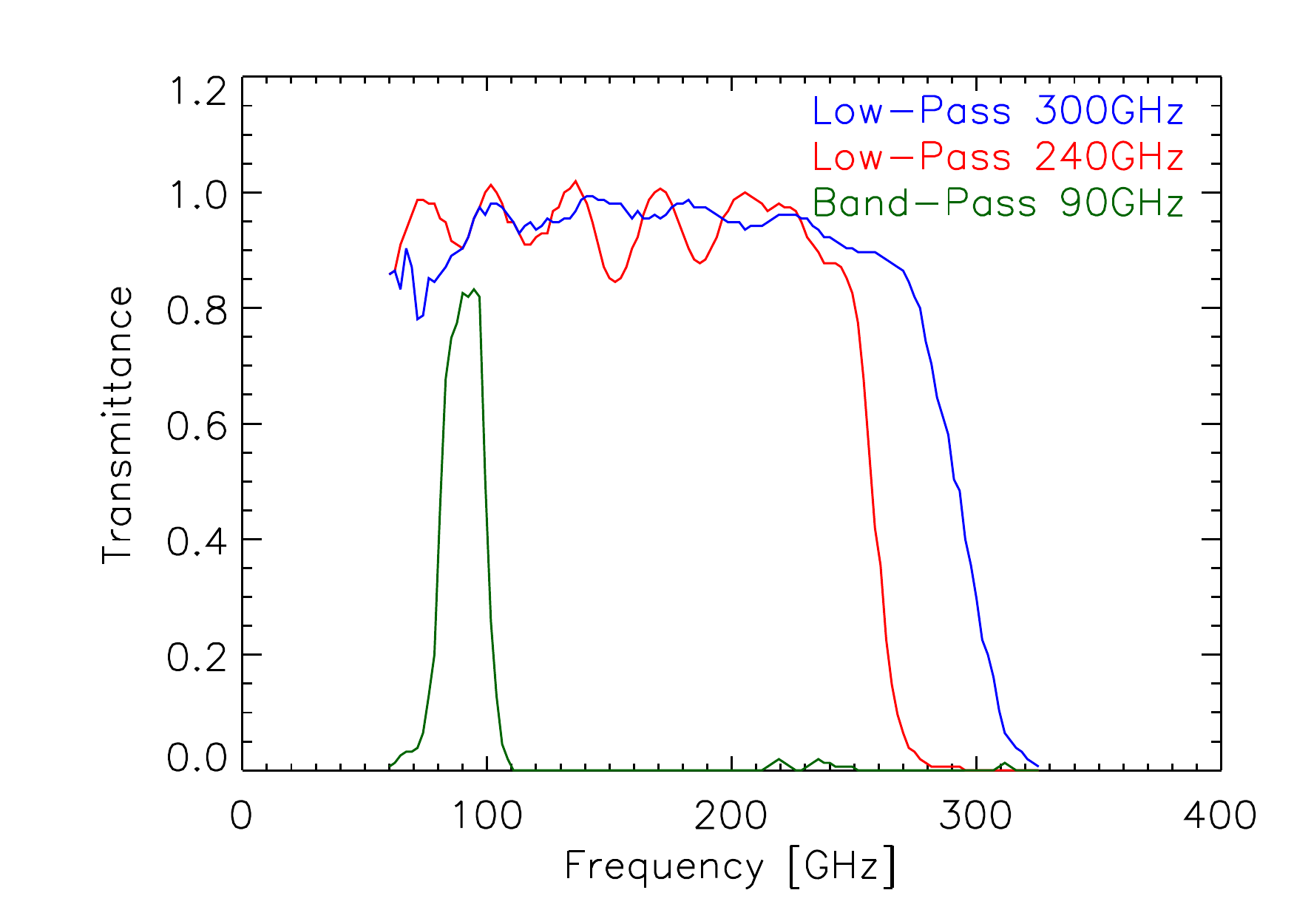}                          
\end{center}
\vspace{-0.5cm}
\caption{The {\it Left} panel is a photo of the cryostat. The {\it Center} panel shows the Zemax design of the optical system. The {\it Right} panel shows the filter transmission spectra. (Color figure online.)}
\vspace{-1cm}
\label{fig:fig3}
\end{figure}

The optical system is composed of a window, two lenses, and a chain of four filters: one thermal shader, two low-pass filters and one band-pass filter; Fig.~\ref{fig:fig3}.

\section{Design}

\begin{figure}[!h]
\begin{center}
\includegraphics[scale=0.29]{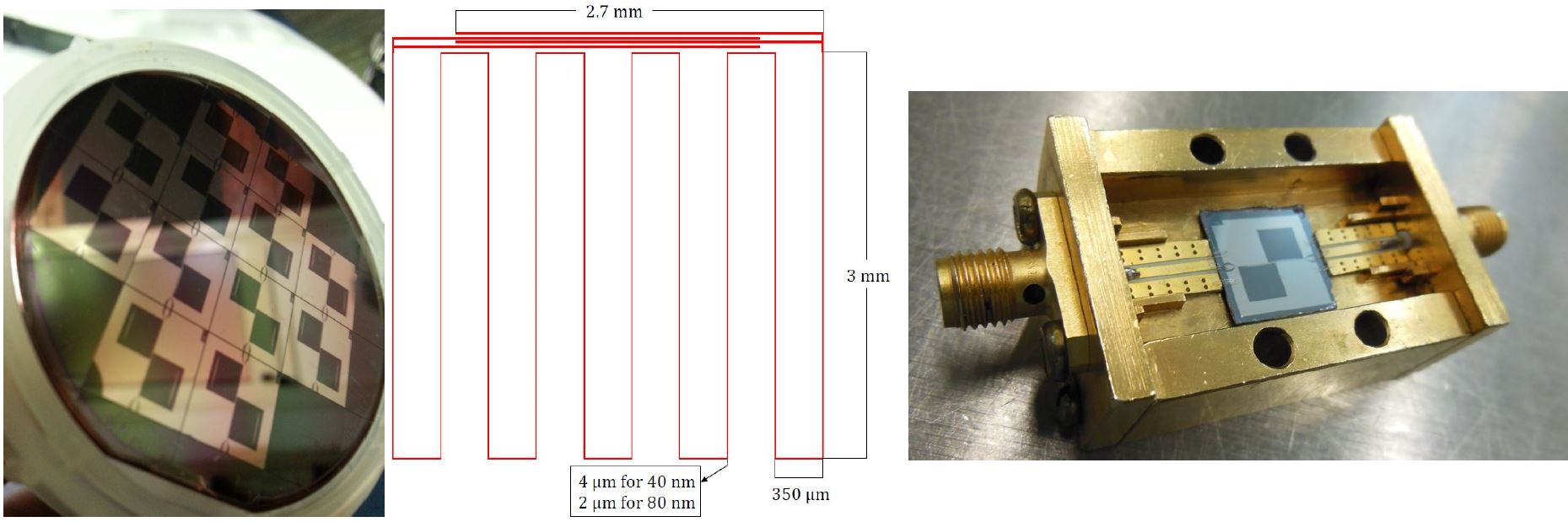}          
\end{center}
\caption{The {\it Left} panel is a photo of the 2" diameter wafer, populated with 12 2-pixel arrays. The {\it Center} panel shows the design of a single pixel. The {\it Right} panel is a photo of 1 array in the holder. (Color figure online.)}
\vspace{-1cm}
\label{fig:fig4}
\end{figure}

We tested three different detector arrays, consisting of three different types of superconductive film: 40 nm thick Al, 80 nm thick Al and $10/25$ nm thick Ti-Al bi-layer. Each array has 2 pixels. The pixel geometry is inspired by the LEKID architecture [6], consisting of a meandered inductor, and a broader interdigited capacitor coupled to a feedline, Fig.~\ref{fig:fig4}.

Our samples are fabricated in a ISO5$-$ISO6 clean room at Consiglio Nazionale delle Ricerche Istituto di Fotonica e Nanotecnologie (CNR-IFN), on high-quality (FZ method) 2"$\times300$ $\mu$m intrinsic Si(100) substrate, with high resistivity ($\rho>10$ k$\Omega$cm) and double side polished [7].

\section{Measurements}

The critical temperature, $T_{c}$, of the film, is linked to the minimum frequency, $\nu_{m}$, able to break CPs by 
\begin{equation}
	h\nu_{m}=2\Delta\left(T_{c}\right)\;. 
\end{equation}
Since the W-band starts at 75 GHz, $T_{c}$ has to be lower than 1.03 K.

We measured $T_{c}$ using the 4-wire reading of the resistance of the feedline at different temperatures, obtaining $T_{c}=\left(1.352\pm0.041\right)$ K and $T_{c}=\left(1.284\pm0.039\right)$ K for Aluminum 40 nm and 80 nm thick, respectively. The uncertainties on the temperature are due to the thermometer calibration, and it is about the 3\% of the measure itself.
   
\begin{figure}[!h]
\begin{center}
\includegraphics[scale=0.44]{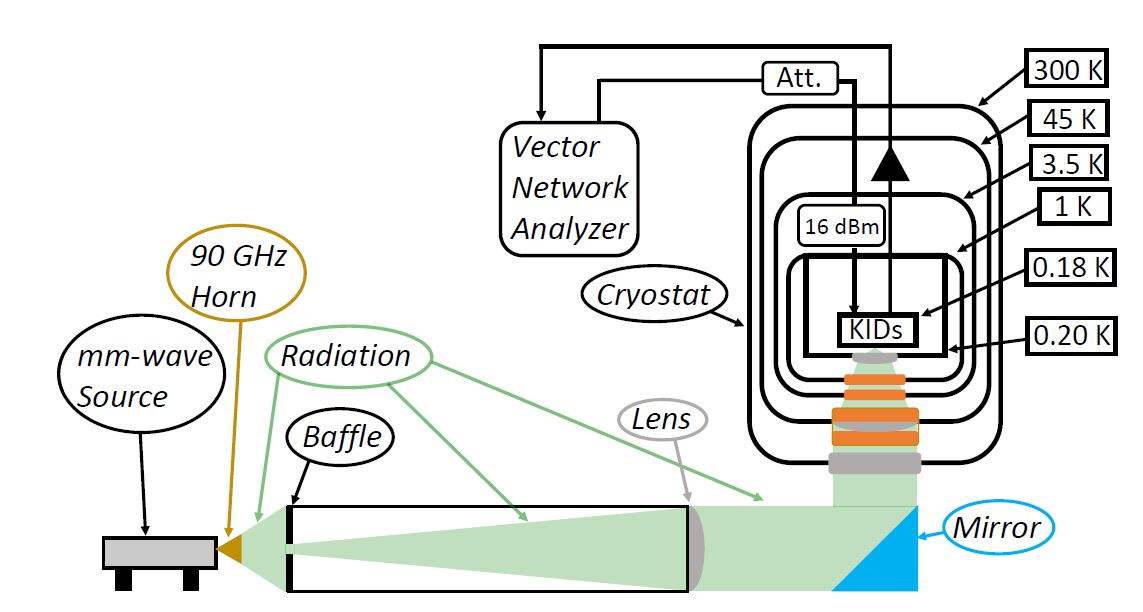}       
\vspace{-0.3cm}
\end{center}
\caption{Setup for the measurement of the critical temperature illuminating the chip with a Millimeter-Wave source (VNA). (Color figure online.)}

\label{fig:fig5}
\end{figure}

\begin{figure}[!h]
\begin{center}
\includegraphics[scale=0.29]{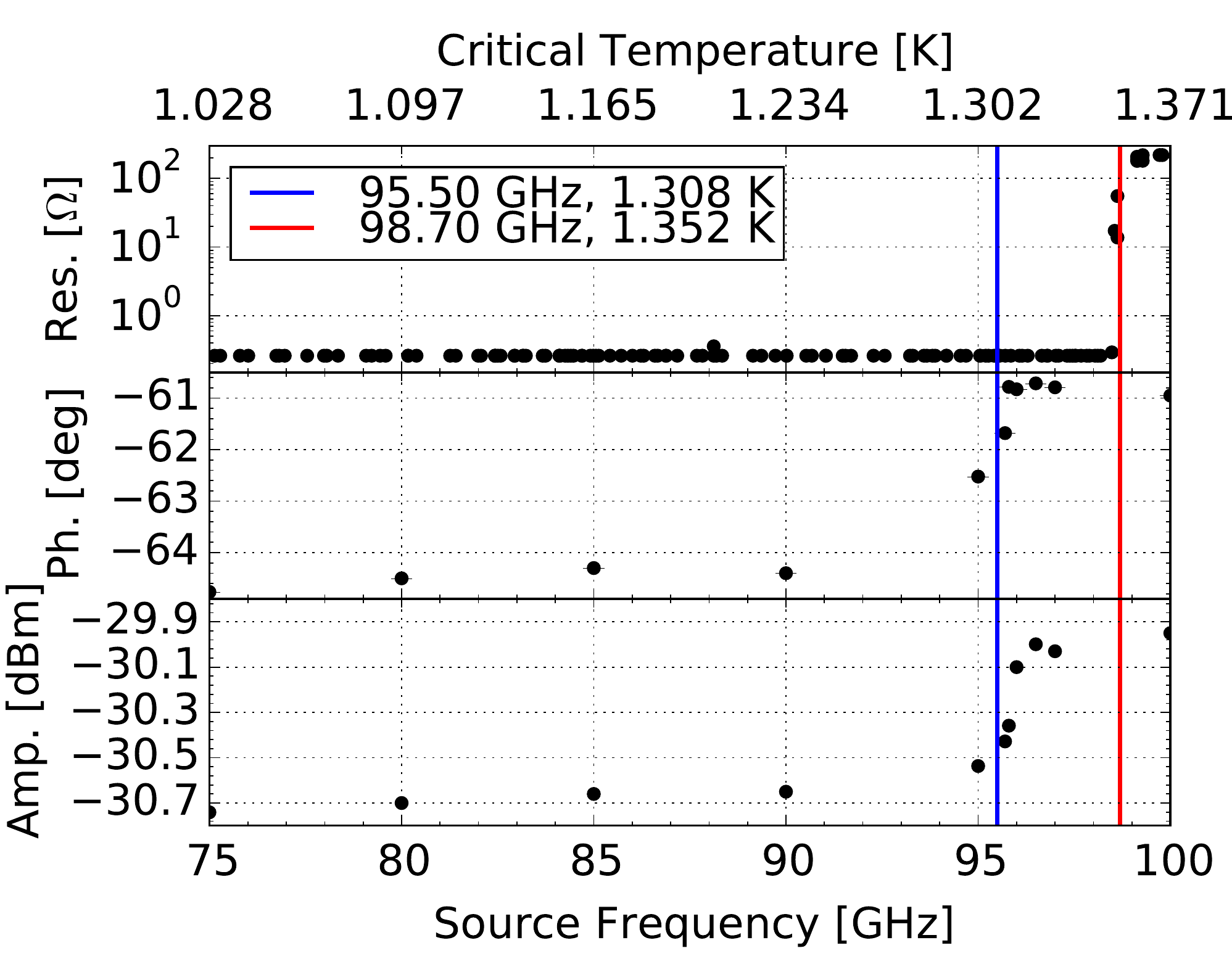}\hspace{-0.17cm}              
\includegraphics[scale=0.29]{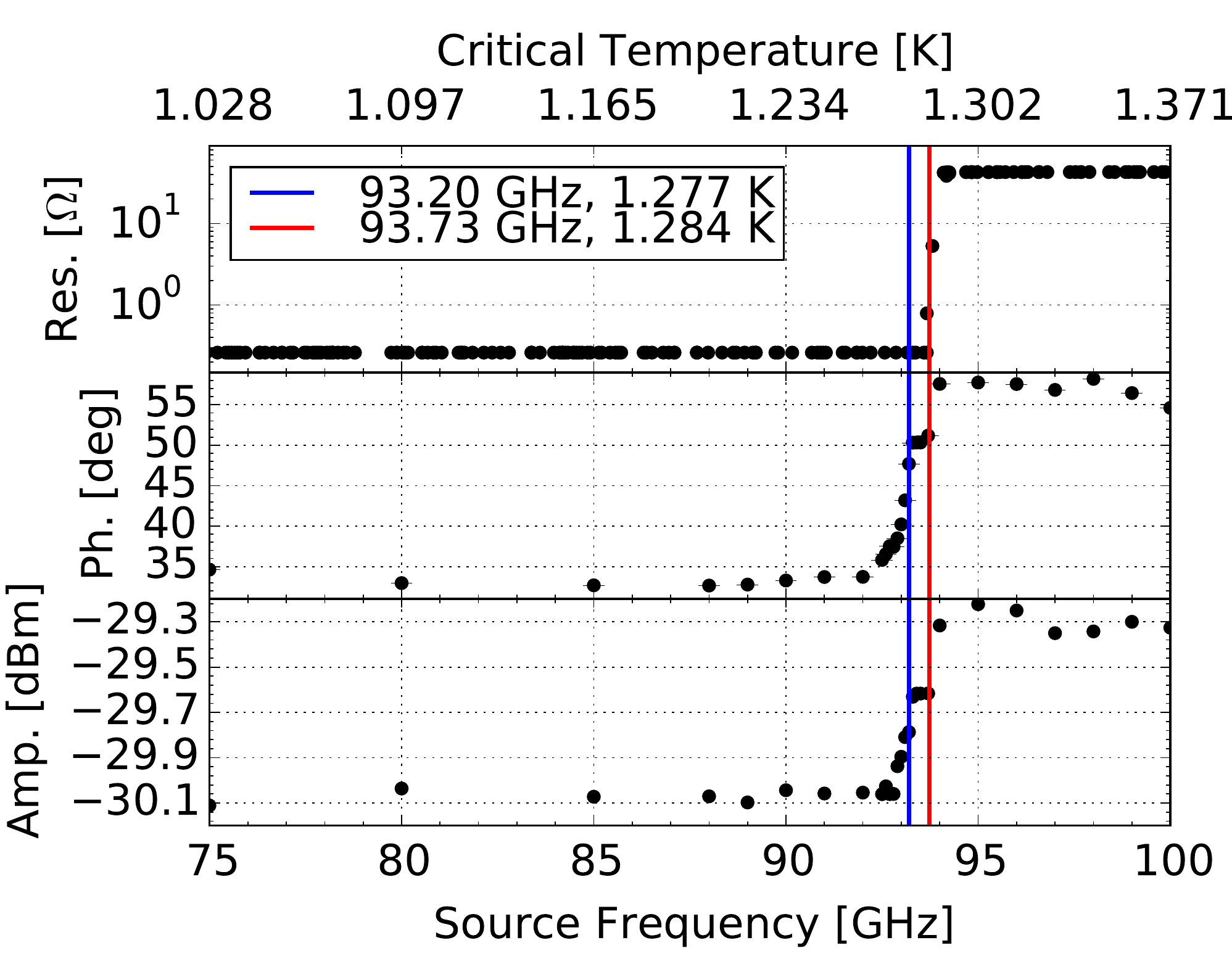}                                        
\end{center}
\vspace{-0.3cm}
\caption{Measurements of the critical temperature for Aluminum 40 nm thick (\emph{Left} panel) and 80 nm thick (\emph{Right} panel). For each panel, the top plot reports the 4-wires resistance measurement result, while the center and bottom plots display the response measurements with a millimeter-wave source (VNA). (Color figure online.)}

\label{fig:fig6}
\end{figure}

Independent evidence of $T_{c}$ was obtained by illuminating the chip with a \\Millimeter-Wave source, Fig.~\ref{fig:fig5}. We monitored phase, amplitude and resonant frequency of the resonators while increasing the frequency of the source from 75 to 110 GHz. The minimum source frequencies producing significant response variations of the resonators are $\nu_{m}=95.50$ GHz and $\nu_{m}=93.20$ GHz for Aluminum 40 nm and 80 nm thick respectively. These values correspond to $T_{c}=1.308$ K and $T_{c}=1.277$ K, in full agreement with the 4-wire measurements, Fig.~\ref{fig:fig6}.

\begin{figure}[!h]
\begin{center}
\includegraphics[scale=0.29]{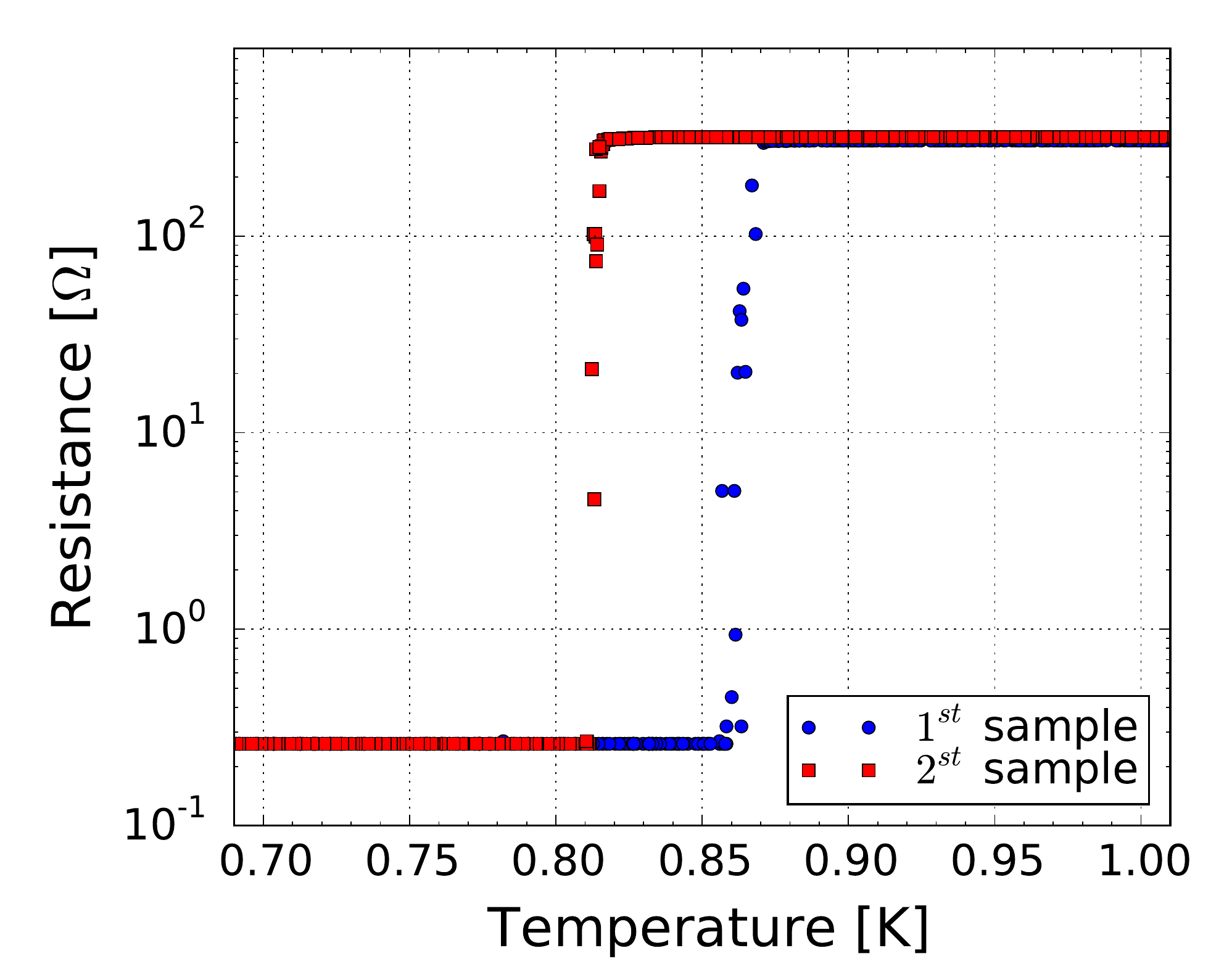}                  
\end{center}
\vspace{-0.3cm}
\caption{Direct measure of the critical temperature of 10 nm thick Ti $+$ 25 nm thick Al bi-layer. (Color figure online.)}

\label{fig:fig7}
\end{figure}

\begin{figure}[!h]
\begin{center}
\includegraphics[scale=0.29]{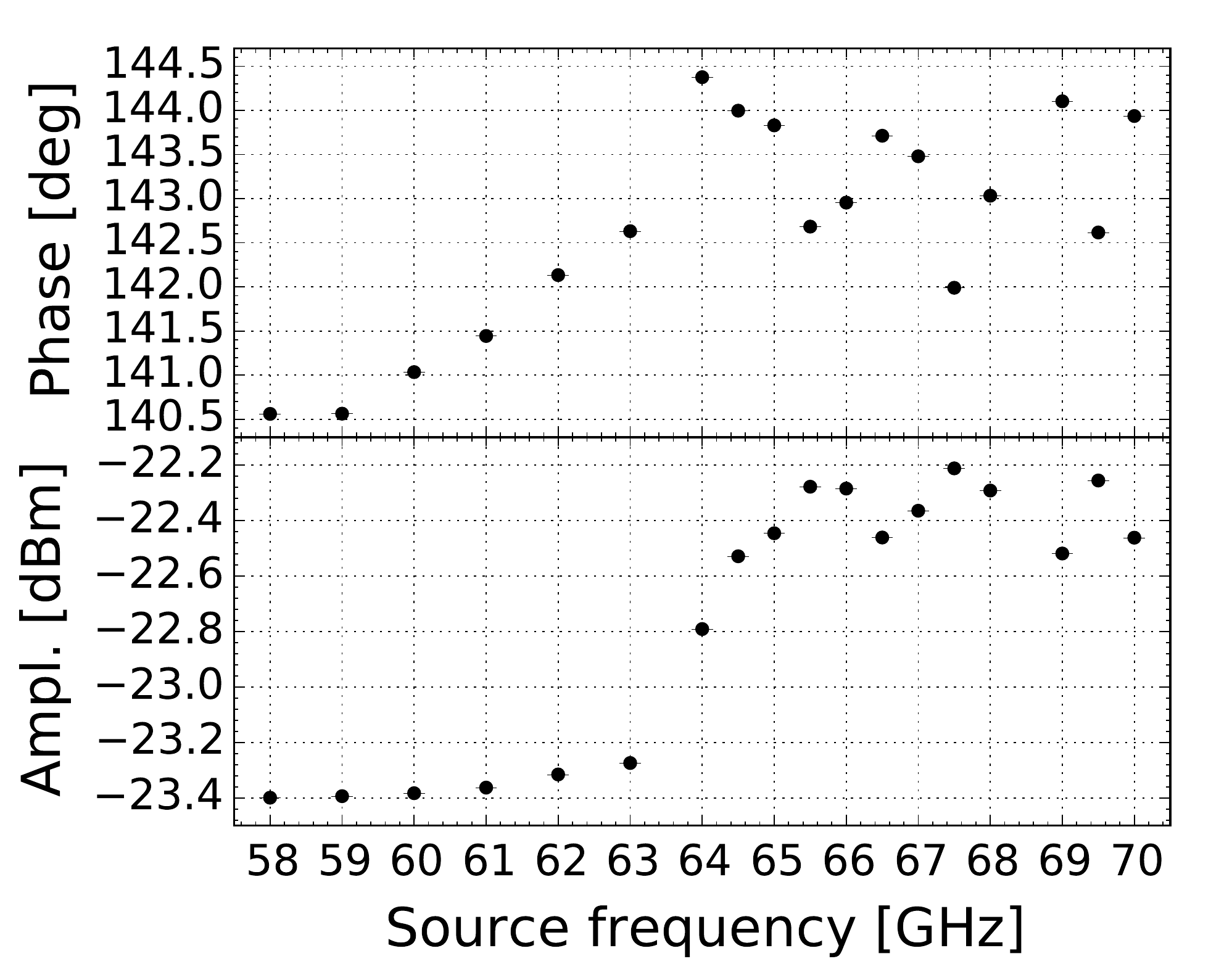}\hspace{-0.2cm}              
\includegraphics[scale=0.29]{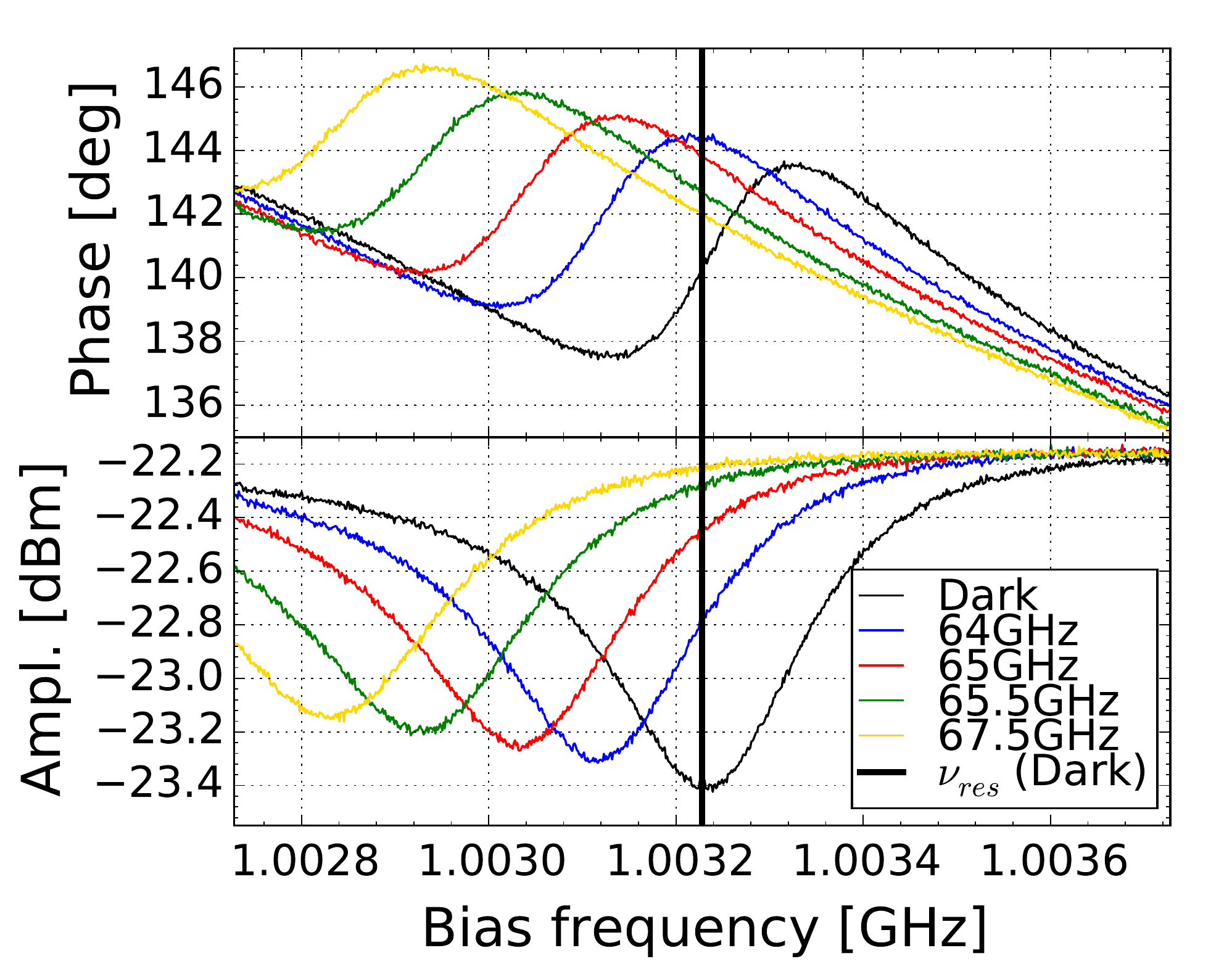}                                        
\end{center}
\vspace{-0.3cm}
\caption{\emph{Left} panel: measurement of the cut-on frequency with a millimeter-wave source (VNA). \emph{Right} panel: resonance responses over 1 MHz span around the dark resonance frequency ($\nu_{res}$), for different source frequencies. $\nu_{res}$ (Dark) is our operation point. (Color figure online.)}

\label{fig:fig8}
\end{figure}

The 80 nm critical temperature is still too high to operate the KIDs in the lowest half of the W-band. In Fig.~\ref{fig:fig7} we show the 4-wire measurement of the $T_{c}$ for two samples of the 10 nm thick Ti $+$ 25 nm thick Al bi-layer: we find $T_c=\left(0.820\pm0.025\right)$ K , which allows for operation of the LEKID in the entire W-band. Using a VNA, we measured the minimum frequency at which the KID sensor responds to incoming radiation. This is about 65 GHz, as shown in Fig.~\ref{fig:fig8} (\emph{Left} panel) where the relative minima in the phase response, like those at 65.5 GHz and 67.5 GHz, are due to the large shift of the resonance frequency (\emph{Right} panel of Fig.~\ref{fig:fig8}). In this measurement setup, a constant excitation power over frequency is warranted by the Signal Generator of the mm-wave source, so the measured response shown in Fig.~\ref{fig:fig8} has to be attributed entirely to the detector.


\section{Conclusion}

Sensitive Detector Arrays working in W-band are very interesting for operation with large radiotelescopes (like the SRT) and in future space missions. Among the targets are the observation of the SZ effect, the surveys of dust and interstellar medium, the study of AGNs and CMB polarization.

For this purpose, we have designed, fabricated, and tested KIDs covering the whole W-band. The use of Ti-Al bi-layers allows extension of the operation of the detector down to frequencies as low as 65 GHz, thus covering the entire W-band.
We are now performing a full characterization of Ti-Al LEKID, including optical and Noise Equivalent Power measurements. The first results confirm promising performance.

\begin{acknowledgements}
We would like to thank Roberto Diana and Anritsu: the measurements between 58 and 70 GHz were made possible thanks to the loan of Anritsu MS4647B VNA.

This research has been supported by Sapienza University of Rome and the Italian Space Agency.
\end{acknowledgements}

\end{document}